# Path integrals and the double slit

Eric Jones, Roger Bach, and Herman Batelaan

*Department of Physics and Astronomy, 208 Jorgensen Hall,*

*University of Nebraska—Lincoln, Lincoln, Nebraska 68588-0299*

Abstract

Basic explanations of the double slit diffraction phenomenon include a description of waves that emanate from two slits and interfere. The locations of the interference minima and maxima are determined by the phase difference of the waves. An optical wave, which has a wavelength $\lambda$ and propagates a distance $L$, accumulates a phase of $2\pi L/\lambda$. A matter wave, also having wavelength $\lambda$ and propagating the same distance $L$, accumulates a phase of $\pi L/\lambda$, which is a factor of two different from the optical case. Nevertheless, the phase difference, $\Delta\varphi$, for interfering matter waves that propagate distances that differ by $\Delta L$, is approximately $2\pi\Delta L/\lambda$, which is the same value computed in the optical case.

The difference between the matter and optical case hinders conceptual explanations of diffraction from two slits based on the matter-optics analogy. In the following article we provide a path integral description for matter waves with a focus on conceptual explanation. A thought experiment is provided to illustrate the validity range of the approximation $\Delta\varphi \approx 2\pi\Delta L/\lambda$.

## I. INTRODUCTION

The presentation of the double slit typically begins with a discussion of Young's original experiments on the diffraction and interference of light.[1] Demonstrations such as a ripple tank, one of Young's own inventions, are used to reinforce the concept of wave interference and Huygens' Principle of the superposition of waves.[2, 3] Figure 1 shows circular waves that impinge on a pair of narrow slits having separation $d$. The slits become themselves new sources of circular waves. The phase associated with a wave is the number of wave-fronts counted along a line with length $L$, that is, $L/\lambda$, multiplied by $2\pi$. Waves interfere constructively when the phase difference is an integer multiple of $2\pi$. These ideas lead to the familiar condition

$$\Delta L = d\sin(\theta) = n\lambda, \quad (1)$$

where $n$ indicates the diffraction order that occurs at the diffraction angle $\theta$. This analysis represented in Fig. 1 describes what we will hereby refer to as the "intuitive approach." Even though this approach is correct for light, it is not for matter. The first problem is that it uses an incorrect phase, $2\pi L/\lambda_{dB}$, for a matter wave (where $\lambda_{dB}$ is the de Broglie wavelength). The second problem is that the use of the intuitive approach will nevertheless give the correct phase difference for most situations.

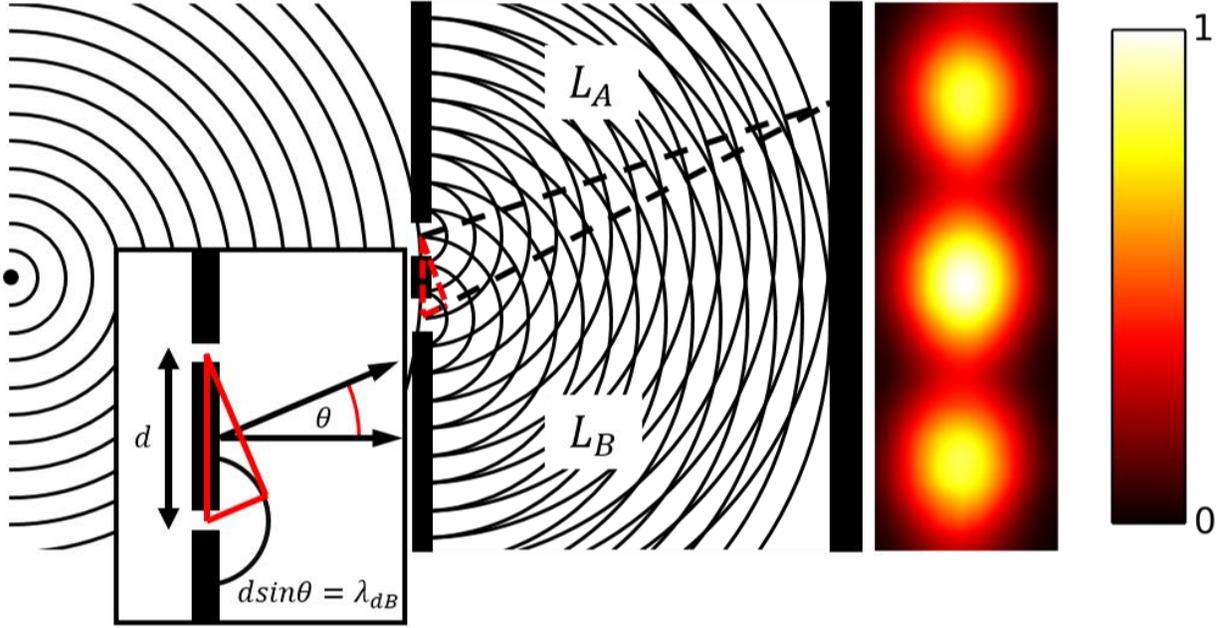

**FIG. 1. Typical schematic of Young's two-slit arrangement. The condition for first order constructive interference, $dsin(\theta) = \lambda$, is illustrated. Shown right is recently published data for an electron double slit interference experiment.**[4] **(Color available online.)**

In this article, the intuitive approach and its limits of validity are discussed for matter waves. The approach is also compared to a stationary phase approach motivated by the path integral formalism. The path integral formalism is shown in Section IIa to give a single path phase of $\pi L/\lambda_{dB}$, and a phase difference between two interfering paths that is approximately $2\pi\Delta L/\lambda_{dB}$. This phase difference agrees with the intuitive approach. The path integral formalism assigns different velocities (and thus wavelengths) to different paths. This appears to be inconsistent with the idea that a double slit is illuminated with a wave described by one velocity or wavelength. This apparent inconsistency will be clarified in the next sections. In Sections IIb, c, and d, the relation to the optical case, wave mechanics, and time-dependence is discussed, respectively. In Sections IIIa through IIIe, a stationary phase argument completes the justification for the path selections made in Section IIa. At this point, it may appear that apart from some conceptual details, the intuitive method's validity can be justified by the path integral method. In section IV, a thought experiment is discussed for which the intuitive approach predicts phase differences that disagree with the path integral method, with the purpose to illustrate that the intuitive method agrees only approximatively.

**IIa. Determining the phase from the path integral**

Feynman developed a method to construct solutions to Schrödinger's equation based on Dirac's observations on the relationship between the evolution of quantum states between points in space-time and the classical notion of particle trajectories.[5, 6, 7] In Feynman's path integral formalism, a probability amplitude is determined from a phase, $\varphi_{path}$, computed along a particular path connecting two space-time events. The total probability amplitude $K(\beta;\alpha)$ for finding a particle at location $x_\beta$ at time $t_\beta$, having started at location $x_\alpha$ at the earlier time $t_\alpha$, is given by the sum

$$K(\beta;\alpha) = \sum_{\text{all paths } \alpha \to \beta} const \cdot e^{i\varphi_{path}}, \qquad (2)$$

where all paths connect events $\alpha$ and $\beta$.[8] The phase $\varphi_{path}$ accumulated along any path is given by

$$\varphi_{path} = \frac{1}{\hbar} \int_{Path} L(x, \dot{x}, t) dt, \qquad (3)$$

where $L$ is the Lagrangian, which depends on the positions, $x$, velocities, $\dot{x}$, and times, $t$, along the path. This path integral approach is used to efficiently describe experimental results for matter interferometry, for example the two slit experiment for electrons.[9, 10]

For the case of a particle moving in a potential $V(x,t)$ in 1-D, the Lagrangian is

$$L(x,\dot{x},t) = \frac{m}{2}\dot{x}^2 - V(x,t). \qquad (4)$$

In the absence of a potential the velocity is constant along the path of integration and the accumulated phase,

$$\varphi = \frac{m}{2\hbar}\left[\frac{x_\beta - x_\alpha}{t_\beta - t_\alpha}\right]^2 (t_\beta - t_\alpha) = \frac{\pi}{\lambda_{dB}}(x_\beta - x_\alpha), \qquad (5)$$

is a function of the endpoints. The phase of a path is thus $\pi L/\lambda_{dB}$.

Now consider a double slit illuminated with a matter wave characterized by one velocity. This system can be characterized by the interference of two paths, represented by the dashed lines of Fig. 1. A reasonable assumption would be that the velocity, and thus the de Broglie wavelength, is the same for both paths. The phase difference between the two indicated paths of lengths $L_A$ and $L_B$ would then be computed from Eq. (5) to be

$$\Delta\varphi = \varphi_B - \varphi_A = \frac{\pi}{\lambda_{dB}}(L_B - L_A). \qquad (6)$$

This result is incompatible with the phase difference expected from the intuitive approach because it differs by a factor of 2. This result is also incompatible with experiment. This is fine because it is indeed incorrect; the false assumption made is that the velocities along both paths are the same. This is not a feature of the path integral formalism. The expected result can be recovered by noting that interfering paths have equal durations $\Delta t$ in time; they both must begin at $\alpha$ and end at $\beta$, as expressed in Eq. (2). Because $L_A$ and $L_B$ are not equal, the consequence is that paths $A$ and $B$ have different velocities. The corresponding de Broglie wavelengths for paths $A$ and $B$ are then

$$\lambda_{A,B} = \frac{h\Delta t}{mL_{A,B}}. \tag{7}$$

The path length difference $\delta L$ between the two paths is taken to be small in comparison to the path length $L_{A,B}$. For $L_A < L_B$, the de Broglie wavelength can be expanded as

$$\lambda_B \cong \frac{h\Delta t}{mL_A}\left(1 - \frac{\delta L}{L_A} + O\left(\left(\frac{\delta L}{L_A}\right)^2\right)\right) \cong \lambda_A\left(1 - \frac{\delta L}{L_A}\right). \tag{8}$$

Terms of order $O\left(\left(\frac{\delta L}{L_A}\right)^2\right)$ are neglected. The phase difference between the two paths is

$$\Delta\varphi = \varphi_B - \varphi_A \cong \frac{\pi L_B}{\lambda_A\left(1 - \frac{\delta L}{L_A}\right)} - \frac{\pi L_A}{\lambda_A} \cong \frac{2\pi(L_B - L_A)}{\lambda_A}. \tag{9}$$

Thus, the correct result is recovered and justified by the path integral formalism of quantum mechanics.

**IIb. Comparison to the optical case.**

The question may arise why the analogous situation of two slit diffraction for light presents no conceptual difficulty. The use of straight paths in Fig. 1 for light could be justified by the application of Fermat's principle of least time.[11] These paths are called rays in the geometrical optics formulation of light propagation.[12,13,14] Rays are constructed from the normals of a succession of electromagnetic wave-fronts. Each ray is associated with a phase called the eikonal $\phi$ that is calculated in a homogeneous medium as

$$\phi = \int_{ray} \mathbf{k} \cdot d\mathbf{l} = \int_{ray} \omega dt. \tag{10}$$

For light, this phase has the value $\int \mathbf{k} \cdot d\mathbf{l} = kL = 2\pi L/\lambda$ along a ray. This justifies the intuitive approach of counting wave-fronts along the propagation paths as in the still pictures of Fig. 1. The equality of the two integrals in Eq. ( 10 ) implies (note $dl/dt = c$) that the dispersion relation for light is linear:

$$\omega = |k|c. \qquad (11)$$

The dispersion relation determines the group and phase velocities. For light propagating in free space, these velocities are the same. Matter wave propagation is different from light because it has a quadratic dispersion relation[15] and the group and phase velocities are not the same. The connection between velocity and phase for matter waves is discussed in the following section.

**IIc. Determining the phase from the wave description: the motion picture.**

Consider the motion of a one dimensional electron wave packet. This superposition of waves $\psi(x,t)$ is given by the Fourier transform of the momentum distribution $f(k - k_0)$ of the constituent waves in the group:

$$\psi(x,t) = \int_{-\infty}^{\infty} f(k - k_0) e^{i(kx - \omega(k)t)} dk. \qquad (12)$$

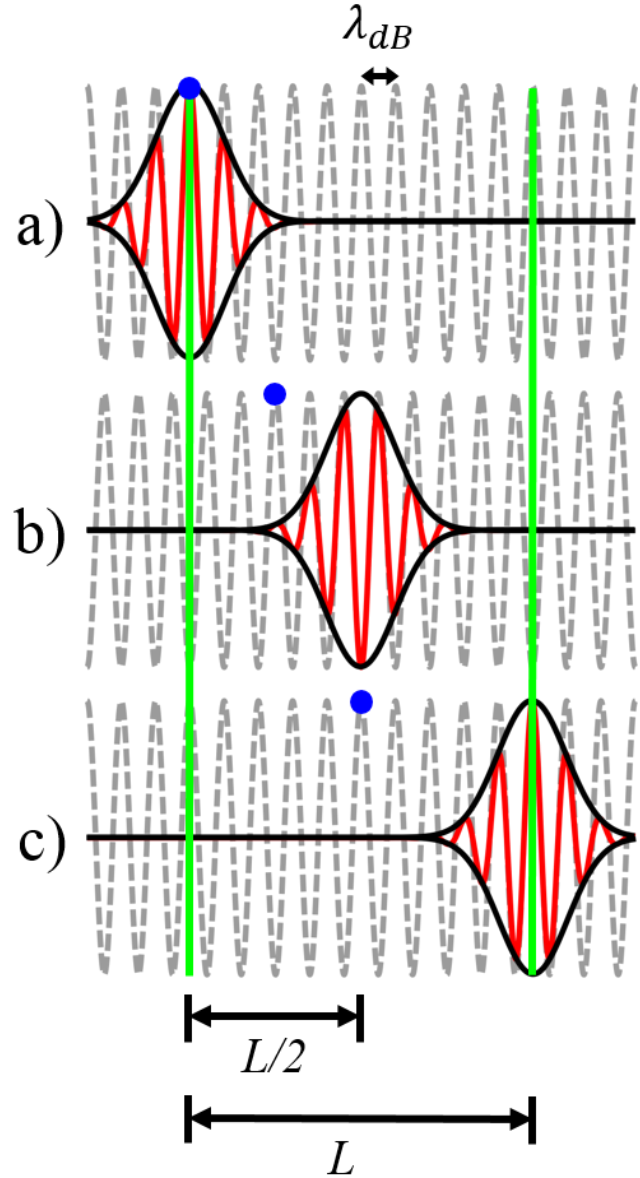

**FIG. 2. Matter wave propagation. Three snapshots of the evolution of a Gaussian wave packet are shown. The carrier wave moves at the phase velocity $v_{ph}$ to the right, indicated with the blue dot. The envelope moves at the group velocity $v_G = 2v_{ph}$. A pulse that propagates a length L accumulates a phase $\varphi = kL - \omega t$. The angular frequency is given by $= kv_{ph}$, while the propagation time is given by $t = L/v_G$. Substitution gives that $\varphi =$**

$\pi L/\lambda_{dB}$, **which differs from the intuitive answer of counting waves along the distance multiplied by $2\pi$.**

For a Gaussian momentum distribution $f(k-k_0) = exp[-(k-k_0)^2/2(\Delta k)^2]$ with width $\Delta k$ and a dispersion relation $\omega(k) = \frac{\hbar k^2}{2m}$, the wave packet $\psi(x,t)$ is then approximately given by

$$\psi(x,t) \propto e^{i(k_0 x - \omega_0 t)} e^{-\frac{(\Delta k)^2}{2}\left(x-\frac{\hbar k_0}{m}t\right)^2} = e^{ik_0\left(x-\frac{\hbar k_0}{2m}t\right)} e^{-\frac{(\Delta k)^2}{2}\left(x-\frac{\hbar k_0}{m}t\right)^2}. \tag{13}$$

A typical matter wave packet spreads as time evolves and the frequency components disperse; however, for sufficiently short times this spreading and dispersion can be neglected. The real part of Eq. ( 13 ) is illustrated in Fig. 2 for three times.[16, 17] The first exponential factor, represented by a dashed gray line in Fig. 2, is a sinusoidal carrier wave travelling with the phase velocity $v_P \equiv \omega/k = \hbar k/2m$. The second factor is the Gaussian envelope, represented by a solid black line, whose center travels twice as fast as the sinusoidal wave at the group velocity

$$v_G \equiv \frac{\partial \omega}{\partial k} = \frac{\hbar k}{m} = 2v_P. \tag{14}$$

The group velocity is the particle velocity and determines the de Broglie wavelength. Suppose that the wave packet in Fig. 2 travels a distance $L$ in a time $\Delta t$. The connection between $L$ and $\Delta t$ is determined by the motion of the center of the Gaussian envelope as

$$L = \frac{\hbar k_0}{m} \Delta t. \tag{15}$$

Substituting this relationship into the phase argument of the carrier wave gives an accumulated phase $\varphi$ of

$$\varphi = \frac{k_0 L}{2} = \frac{\pi L}{\lambda_0}. \qquad (16)$$

Thus the phase accumulated by a matter wave packet moving from one position to another follows from inspecting a time-dependent solution.

**IId. The time-independent and time-dependent Schrödinger equations**

The fact that none of the experimental parts of a double slit experiment change over time suggests inspecting a steady state solution. Consider the time-dependent Schrödinger equation,

$$\frac{-\hbar^2}{2m} \nabla^2 \psi - V\psi = i\hbar \frac{\partial \psi}{\partial t}. \qquad (17)$$

When the potential $V$ in the Schrödinger equation does not depend on time, then the time-independent equation is derived from the time-dependent equation by separation of variables and division by the common factor $e^{-i\omega t}$. This results in the time-independent Schrödinger equation,

$$\nabla^2 \varphi + \frac{2(E-V)}{m\hbar^2} \varphi = 0. \qquad (18)$$

The factor $2(E-V)/m\hbar^2$ can be defined as $k^2$ to give the Helmholtz equation,

$$\nabla^2 \varphi + k^2 \varphi = 0, \qquad (19)$$

the solutions of which also describe the steady state solutions for optics. This well-known analogy is a defining property of the field of matter-optics.[18]

Solutions to the Helmholtz equation with the same energy values (and thus k-values for free space solutions) can be summed to construct a new solution and the superposition principle holds. The circles in Fig. 1 can then be thought of as depicting the wave fronts of the real part of the circular waves $\varphi_A(r_A)$ and $\varphi_B(r_B)$ that are solutions to the Helmholtz equation. The probability to find an electron at a location $r$ is given by the Born rule, $|\psi(r,t)|^2 = |\varphi_A(r) + \varphi_B(r)|^2$. The result is time-independent because the time-dependent factor $e^{-i\omega t}$ factors out of the wave function. The condition $\Delta L = n\lambda$ to find maxima in the probability distribution now follows directly. This paper, however, is not focused on this time-independent description, but on the time-dependent description and finding the relation $\Delta\varphi = 2\pi n$, which requires us to set up a relation between length $L$ and time-dependent phase $\varphi$.

**IIIa. Revisiting the path integral propagator in free space**

In Section IIa, we restricted the discussion of path integral phase differences to a particular selection of two paths. However, the full path integral description for the double slit calls for a summation of probability amplitudes over *all* possible paths, not just the selection. First, we will consider 1-D paths in free space as a step towards the double slit case in 2-D. The 1-D propagator will inform the correct choices of paths for the double slit.

The probability amplitude for a particle to travel in free space from space-time event $\alpha$, denoted $(x_0, t_0)$, to event $\beta$, denoted $(x_N, t_N)$, in a number $N$ evenly-spaced time intervals $\epsilon$, is given in Eq. (3.2) of Ref. 8 as

$$K(\beta;\alpha) = \lim_{\epsilon \to 0} \left(\frac{m}{2\pi i \hbar \epsilon}\right)^{N/2} \int \ldots \int \exp \frac{im}{2\hbar\epsilon} \left\{\sum_{j=1}^{N} (x_j - x_{j-1})^2\right\} dx_1 \ldots dx_{N-1}. \tag{20}$$

Feynman points out that the resulting nested Gaussian integrals can be performed iteratively, leading to the result

$$K(\beta;\alpha) = \left(\frac{m}{2\pi i \hbar \cdot N\epsilon}\right)^{1/2} \exp\left\{\frac{im}{2\hbar \cdot N\epsilon}(x_N - x_0)^2\right\}. \tag{21}$$

When the subscripts *0* and *N* are associated with their space-time events $\alpha$ and $\beta$, and the total time $N\epsilon$ is replaced with the time difference, $t_\beta - t_\alpha$, then Eq. ( 21 ) becomes

$$K(\beta;\alpha) = \left(\frac{m}{2\pi i\hbar \cdot (t_\beta - t_\alpha)}\right)^{1/2} \exp\left\{\frac{im}{2\hbar \cdot (t_\beta - t_\alpha)}(x_\beta - x_\alpha)^2\right\}. \tag{22}$$

This result, which is readily generalized to higher dimensions, shows that the amplitude associated with the summation over all possible paths connecting two events in free space is equal to the amplitude associated with the classical path alone, as sketched in Fig. 3.

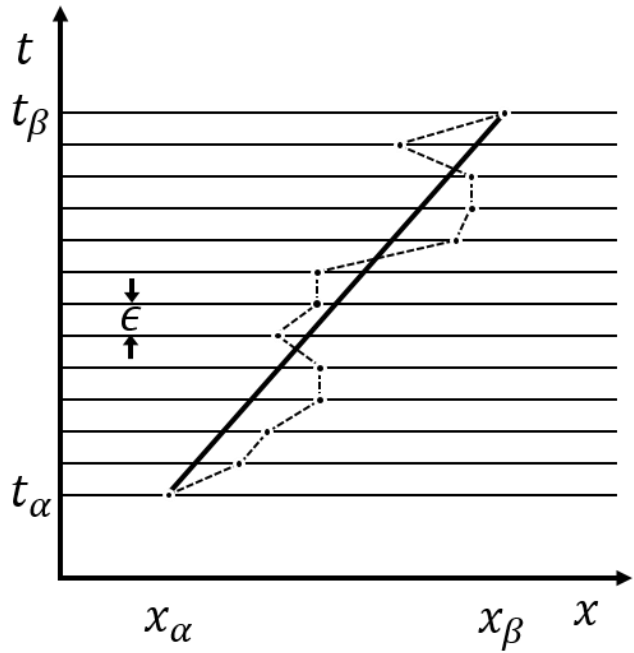

**FIG. 3. Feynman paths for 1-D free particle propagation. A general path (dotted line) is shown for a point particle that travels from space-time point $(x_\alpha, t_\alpha)$ to $(x_\beta, t_\beta)$. The locations *x* that the path crosses (indicated for times separated by $\epsilon$) can by varied along the x-axis. The classical path is indicated with the bold dark line. Feynman showed that the total amplitude for motion from $\alpha$ to $\beta$ summed over all paths (by integrating over the x-**

locations) **is identical to the amplitude computed along the classical path alone, which is a central result from the path integral formalism.**[5, 8]

### IIIb. Two-step propagator for a slit

As the next step towards describing the double slit, consider the amplitude for an electron path intersecting a single slit. This path is described by three space-time events, labelled as in IIIa: the source, defined as the event $\alpha$; the slit crossing, denoted by $slit$; and the measurement at the screen, $\beta$.

The amplitude for such a path can be constructed as the product of the amplitude for two steps. The first step is to reach the slit from $\alpha$, and the second step is to travel from the slit to $\beta$.[19, 20] Applying the result of Eq. ( 22 ) to this case, one obtains

$$K(\beta; \alpha) = K(\beta; slit) \cdot K(slit; \alpha)$$

$$= \left(\frac{m}{2\pi i \hbar \cdot (t_\beta - t_{slit})}\right)^{1/2} exp\left\{\frac{im}{2\hbar \cdot (t_\beta - t_{slit})}(x_\beta - x_{slit})^2\right\}$$

$$\cdot \left(\frac{m}{2\pi i \hbar \cdot (t_{slit} - t_\alpha)}\right)^{1/2} exp\left\{\frac{im}{2\hbar \cdot (t_{slit} - t_\alpha)}(x_{slit} - x_\alpha)^2\right\}$$

$$= \frac{m}{2\pi i \hbar} \cdot \left(\frac{1}{(t_\beta - t_{slit}) \cdot (t_{slit} - t_\alpha)}\right)^{1/2}$$

$$\cdot exp\left\{\frac{imL^2 \cdot (t_\beta - t_\alpha)}{2\hbar \cdot (t_\beta - t_{slit}) \cdot (t_{slit} - t_\alpha)}\right\},$$

( 23 )

where the substitution $(x_\beta - x_\alpha)/2 = L$ was made. We note that this is an approximation: an exact construction for the propagator would take into account the boundary conditions set by the walls. The integrations from $-\infty$ to $\infty$ in Eq. ( 20 ) include paths that pass through the walls; therefore, the propagator in Eq. ( 23 ) adds extraneous paths to the sum. The times $t_\alpha$ and $t_\beta$ defining the boundaries of this path are fixed, but the slit-crossing time, $t_{slit}$, is not. It is not a measured event in the same sense as $\alpha$ or $\beta$ and thus cannot be specified. The total amplitude to cross the slit, $K(\beta; \alpha)$, is then a sum over all of the amplitudes having every possible value of $t_{slit}$.

### IIIc. Time summed amplitude for two-step propagator

To obtain the total amplitude to cross the slit, $K(\beta; \alpha)$, consider the sum of the products $K(\beta; slit) \cdot K(slit; \alpha)$ of Eq. ( 23 ) for every value of $t_{slit}$ occurring between $t_\alpha$ and $t_\beta$. The result is written as

$$K(\beta; \alpha) = \sum_{t_{slit}=t_\alpha}^{t_\beta} K(x_\beta, t_\beta; x_{slit}, t_{slit}) \cdot K(x_{slit}, t_{slit}; x_\alpha, t_\alpha). \qquad (24)$$

For more detail, see the derivation of Eq. ( 24 ) in Appendix A. This formally establishes the sum over intermediate times that is required from the sum over all paths given in Eq. ( 20 ). The sum in Eq. ( 24 ) over the continuous value $t_{slit}$ is proportional to the integral $I(\beta; \alpha)$, given by

$$I(\beta; \alpha) = \int_{-\frac{t_\beta}{2}}^{\frac{t_\beta}{2}} dt \, \frac{m}{2\pi i \hbar} \cdot \left( \frac{1}{\left(\frac{t_\beta}{2} - t\right) \cdot \left(\frac{t_\beta}{2} + t\right)} \right)^{1/2} \cdot \exp\left\{ \frac{imL^2 \cdot t_\beta}{2\hbar \cdot \left(\frac{t_\beta}{2} - t\right) \cdot \left(\frac{t_\beta}{2} + t\right)} \right\}. \qquad (25)$$

Here, $t_\alpha = 0$, $(x_\beta - x_\alpha)/2 = L$ as before, and the variable time $t$ at the slit has been defined to give a symmetric integrand. This integral is derived and evaluated in greater detail in Appendix B. The result, given in terms of the complementary error function, $erfc(z)$,[21] is

$$I(\beta; \alpha) = \frac{m}{2\pi i\hbar} \pi \cdot erfc(-i\sqrt{i\varphi_0}) \qquad (26)$$

with $\varphi_0 \equiv 2mL^2/\hbar t_\beta$. Equation ( 26 ) has an asymptotic expansion given in Eq. 7.1.23 of Ref. 21 as

$$I(\beta; \alpha) \approx \frac{m}{2\pi i\hbar} \sqrt{\frac{\pi}{\varphi_0}} e^{i\varphi_0} e^{i\frac{\pi}{4}} \left[1 - \frac{i}{2\varphi_0} - \frac{3}{4\varphi_0^2} + \frac{15i}{8\varphi_0^3} + \cdots \right]. \qquad (27)$$

The form of Eq. ( 27 ) illustrates that the total integrated amplitude experiences a phase shift of $\pi/4$ from the phase $\varphi_0$. This phase difference is independent of the choice of path and thus the global phase $\pi/4$ can be factored out. Figure 4a shows the convergence of the real part of the numerical evaluation of Eq. ( 25 ) (blue curve) to the real part of the analytic result of Eq. ( 26 ) (black dashed), for $L = 3.37 \times 10^{-6}$ m and $t_\beta/2 = 3.36 \times 10^{-13}$ s. The numeric results are computed for variable limits of integration and plotted as a function of the total time interval being integrated. Figure 4b gives the phase argument of the numeric results (red curve) to show the convergence of the rotation from the initial phase argument given by $\varphi_0$, which is defined by choice of the parameters to be $-\pi$ (black dotted), to an angle of $-3\pi/4$ (black dashed). This establishes the appropriate choice of amplitude for a path crossing one slit. The physical meaning of $\varphi_0$ will now be discussed.

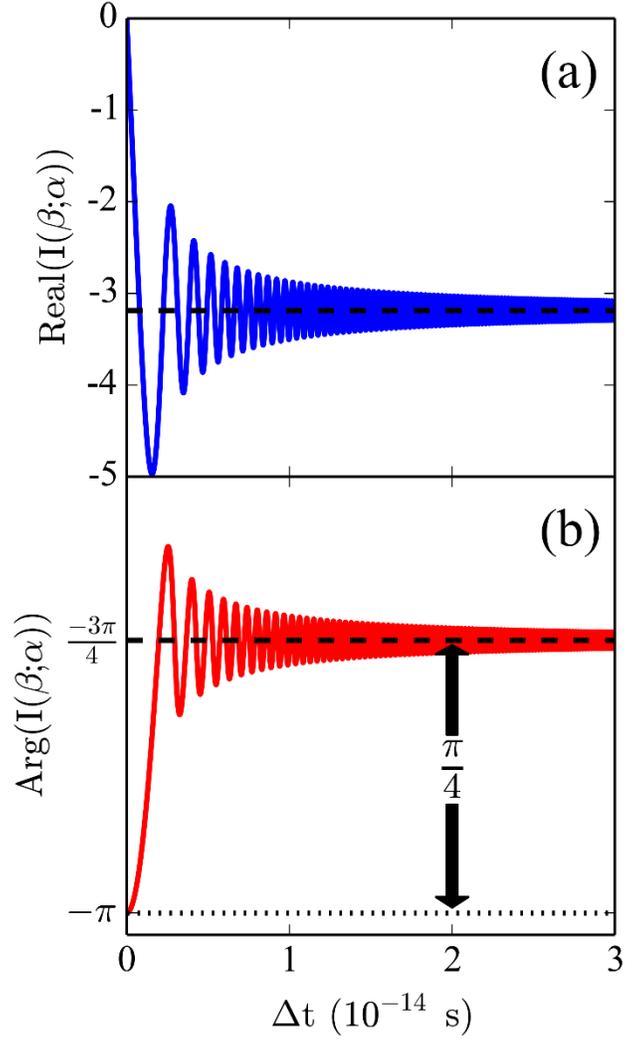

**FIG. 4. Numerical time integration for a single electron path. (a) Real part of Eq. ( 25 ) (solid blue), integrated from $-\Delta t/2$ to $\Delta t/2$ relative to $t_\beta/2$, showing convergence to the analytic value given in Eq. ( 26 ) (dashed line). (b) Complex argument of the integrated amplitude (solid red), showing convergence to $\pi/4$ phase shift (dashed line) from the argument of $e^{i\varphi_0}$ (dotted black). The amplitude proportional to $e^{i\varphi_0}$ is therefore the appropriate choice of a single amplitude to characterize the entire sum.**

### IIId. Stationary phase for the two step propagator

The phase $\varphi$ accumulated along a path crossing a slit is determined from Eq. ( 5 ) to be

$$\varphi = \frac{m}{2\hbar}\left(\frac{L_1^2}{t_{slit}} + \frac{L_2^2}{\tau - t_{slit}}\right), \qquad (28)$$

where $\tau = t_\beta - t_\alpha$, $L_1$ is the path length from source to slit, and $L_2$ is the length from slit to screen. When $t_{slit}$ is varied by $\delta t_{slit}$, the phase can be expanded as a power series in $\delta t_{slit}$ as

$$\varphi = \varphi_0 + \frac{\partial \varphi}{\partial t_{slit}}\delta t_{slit} + \frac{1}{2}\frac{\partial^2 \varphi}{\partial t_{slit}^2}(\delta t_{slit})^2 + \cdots, \qquad (29)$$

where $\varphi_0$ is associated with a particular choice of $t_{slit}$. The first-order term of Eq. ( 29 ) is written out

$$\frac{\partial \varphi}{\partial t_{slit}} = \frac{m}{2\hbar}\left(\frac{L_2^2}{(\tau - t_{slit})^2} - \frac{L_1^2}{t_{slit}^2}\right). \qquad (30)$$

The factor $\partial\varphi/\partial t_{slit} = 0$ when $L_2/(\tau - t_{slit}) = L_1/t_{slit}$: that is, when the velocities along the path are equal before and after the slit. The phase $\varphi$ will then experience no first-order variation from $\varphi_0$ when $t_{slit}$ is chosen by this condition. We then say that the phase is stationary for this choice of path, and the value of the stationary phase is $\varphi_0$. As shown in Section IIIc, this phase characterizes the amplitude arising from the sum of choosing all values of $t_{slit}$; therefore, it is the appropriate choice for a single path. The phase in terms of the de Broglie wavelengths along this single path is now $\pi L_1/\lambda_{dB} + \pi L_2/\lambda_{dB} = \pi L_{path}/\lambda_{dB}$.

### IIIe. Stationary phase in the double slit

Figure 5a shows two interfering paths (green and red) in a space-time diagram for the double slit. The times $t_{slit1}$ and $t_{slit2}$, when paths 1 and 2 intersect the slits, respectively, take any value between the initial time $t_{initial}$ and final time $t_{final}$. The probability distribution at the screen is shown to the right of the screen as an intensity plot. In Fig. 5b, a phasor diagram of the complex amplitudes for the varying times $t_{slit1}$ and $t_{slit2}$ is shown. The highlighted paths in Figs. 5a and 5b are the paths of stationary phase. In Fig. 5c, the phases corresponding to the amplitudes in (b) are given as a function of time to illustrate the stationary phase behavior.

Notice that the stationary phase time for path 1, indicated by the largest red dot in Fig. 5a, occurs after the stationary phase time for path 2. The reason is that the length of path 1 (that is, the length of the dashed line in the x-y plane) is shorter than the length of path 2. As the initial time and final times are the same for both paths, the velocities of the paths are different. The equal length of the part of both paths between the source and slits explains the difference in the stationary phase times for this example. For some other path integral calculations, the slit crossing times are chosen to be identical for all paths,[10, 19] while for the intuitive method, the times are the same for paths of the same length from source to slit.

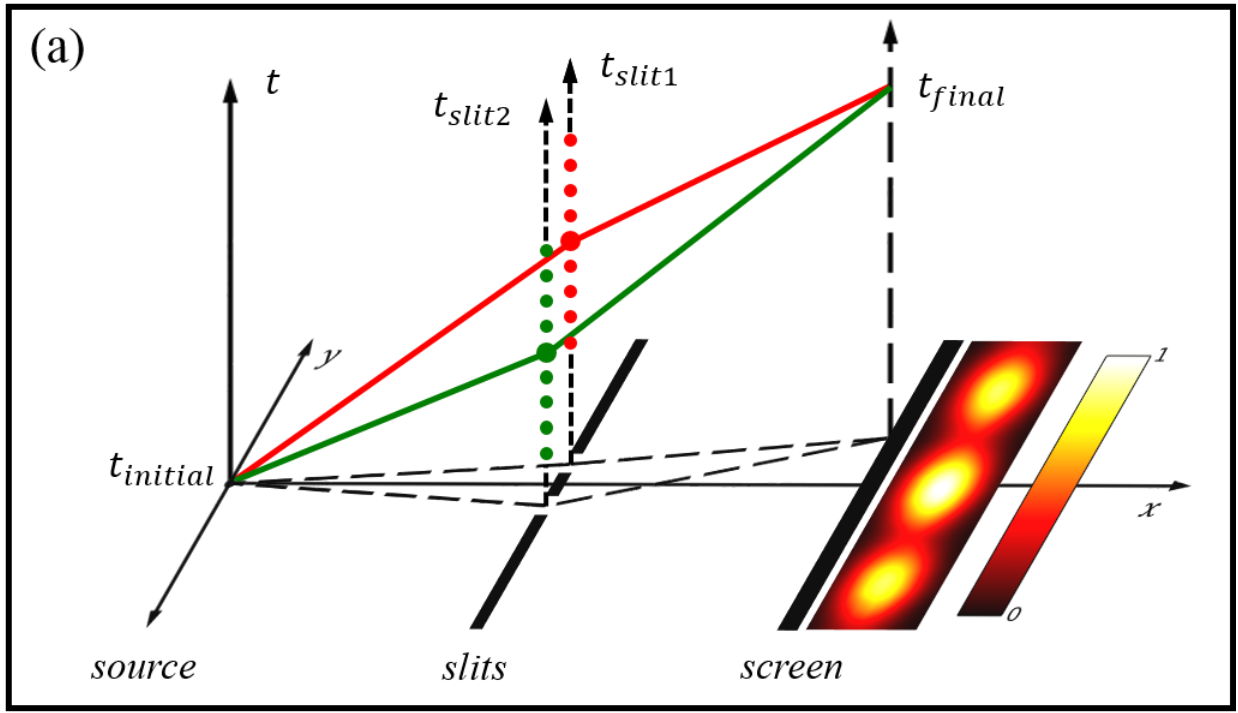

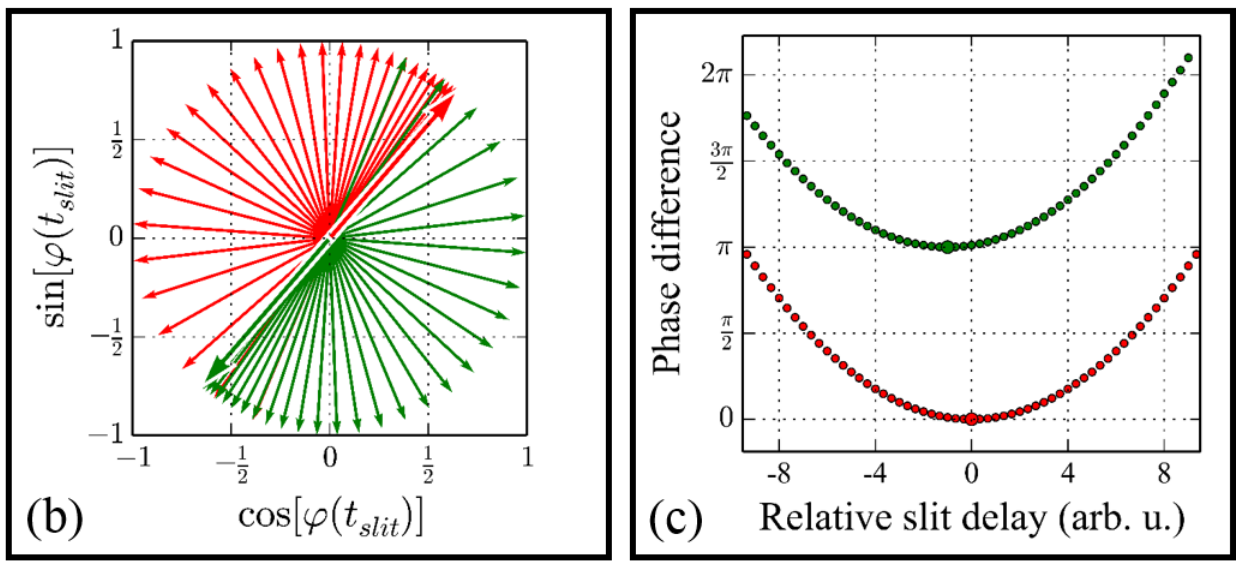

**FIG. 5. Path integral illustration for destructive interference for the two slit arrangement.** (a) The times $t_{slit1}$ and $t_{slit2}$ at which the paths intersect the slits take any value between the initial time $t_{initial}$ and final time $t_{final}$. The resulting probability distribution at the screen is the square of the sum of the amplitudes for all of the times $t_{slit1}$ and $t_{slit2}$. (b) Shown is a phasor diagram for complex amplitudes associated with the intermediate times

for slit 1 (red) and slit 2 (green). The highlighted paths in (b) are the paths of stationary phase shown in (a). (c) The phases corresponding to the amplitudes in (b) are shown as a function of intermediate time to illustrate the stationary phase behavior.

## IV. Phase matters

Do the intuitive method and the path integral method make the same predictions? In other words: "Does the phase of a single path (Eq. ( 5 )) matter?" After all, real experiments are only sensitive to phase differences, which were shown to agree for the intuitive approach and the path integral formalism in Eq. ( 9 ). This agreement is not always the case. Consider now the double slit arrangement in Fig. 6, where the electron source and observation point are in-line with one of the slits. Computing the phases of the drawn paths by Eq. ( 5 ) (dashed lines) leads to a phase difference

$$\Delta\varphi_{path\ integral} = \frac{2md^2}{\hbar\Delta t}. \qquad (31)$$

This result does not depend on the length $L$ in this configuration. If instead we use the intuitive method, we compute the phase difference

$$\Delta\varphi_{intuitive} = 2\frac{2\pi}{\lambda_{dB}(v)}\left(\sqrt{L^2 + d^2} - L\right) \cong \frac{2md^2}{\hbar\Delta t} - \frac{md^4}{2\hbar\Delta t L^2}, \qquad (32)$$

where the time and velocity between the two methods are connected by $v = 2L/\Delta t$. The phase difference now depends on $L$. The phase difference thus depends on the choice of method used for single path phases. When $d^2/4L^2 \sim \pi\hbar\Delta t/2md^2$, Eq. ( 32 ) conflicts with Eq. ( 31 ), and the choice of method matters.

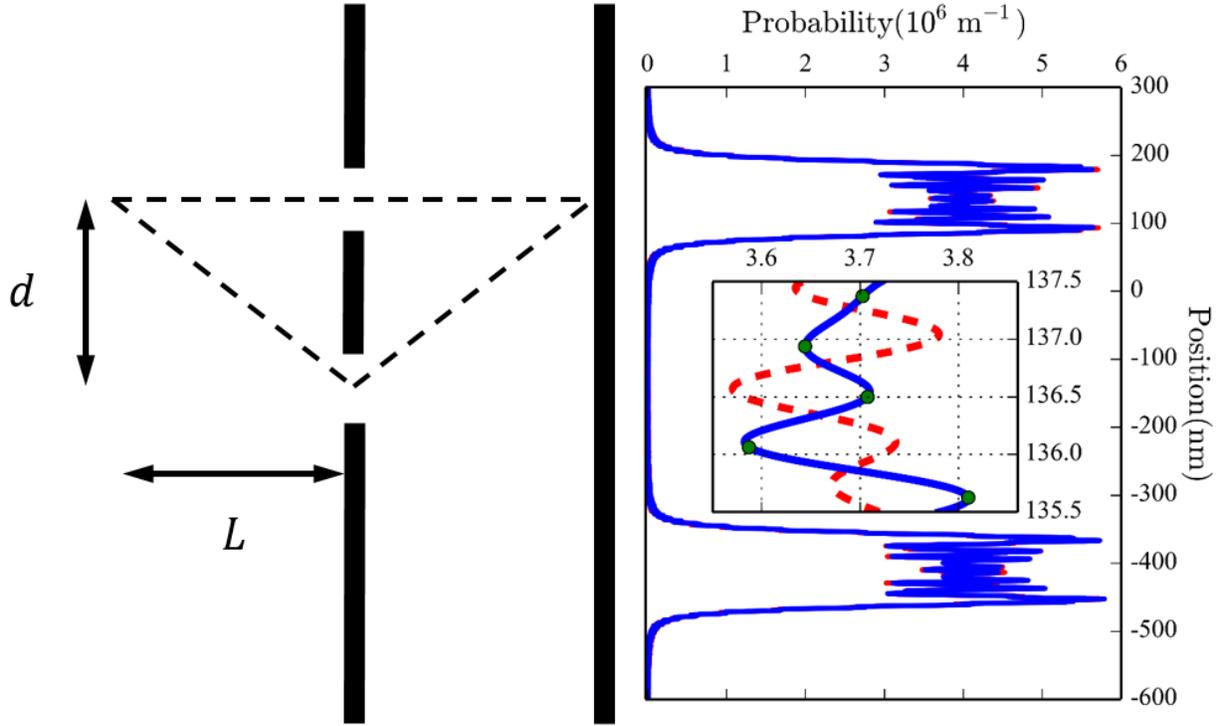

**FIG. 6. Feynman paths and probability distribution in a near-field two slit arrangement. The source is positioned in line with one of the slits and the detection point. The slit separation *d*, propagation length *L*, and velocity are chosen to highlight the discrepancy between the predictions of the two methods. The normalized probability distribution functions at the screen are computed with the path integral stationary phases (blue), the time-summed amplitudes (green points), and the intuitive method (dashed red).**

To best exemplify this conflict, let's now choose the experimental conditions so that the common term of Eqs. ( 31 ) and ( 32 ) is set to an integer multiple of $2\pi$, and the second term of Eq. ( 32 ) set to $\pi$, so that Eq. ( 31 ) predicts constructive interference, while Eq. ( 32 ) predicts destructive interference. For electron diffraction in the symmetric double slit arrangement of Fig. 6, a slit separation of 273 nm with widths of 63 nm can be chosen. Note that such a double slit has been demonstrated recently for electron diffraction in Ref. 4. In contrast to Ref. 4, now the source and screen are placed at the much closer distances of 3.37 $\mu$m from the slits. When $\Delta t$ is fixed for the path integral method by choosing the electron velocity to be $10^7$ m/s over the straight path, the difference of $\pi$ is set between the predictions of the two methods. The diffraction patterns are computed with both methods and shown on the right of Fig. 6. At a location on the detection screen that is in line with the source (at y = 136.5 nm), the path integral method (solid blue) predicts a constructive maximum, while the intuitive method (dashed red) predicts a minimum. A time sum of the form of Eq. ( 24 ) performed for five points on the observation screen over intervals of $6.88 \times 10^{-14}$ s centered on the stationary phase time of each point (green points) agrees with the blue curve computed with the stationary phase times alone. This configuration is experimentally challenging to realize. Nevertheless, near field interferometry for matter waves does exist and may be pushed towards this regime.[22, 23] In conclusion, phase difference predictions from the intuitive method and the path integral formalism will not agree in some near-field conditions. While the global phase of a single path does not matter, to obtain correct phase differences, the single path phases must be handled appropriately.

## V. Summary and conclusions

The intuitive approach can give excellent approximate phase differences in most situations and thus leads to the correct prediction of the positions of interference extrema. This approach is justified by considering stationary solutions to the Schrödinger equation. The conceptual trap is that a student may infer from the correct phase difference, $2\pi\Delta L/\lambda$, that the phase of a single path is given by $2\pi L/\lambda$ (as would be correct for optical waves). The path integral description of quantum mechanics gives the correct phase difference $2\pi\Delta L/\lambda$ between paths, the correct phase $\pi L/\lambda$ accumulated over time along a single path, and justifies drawing "paths" in space to compute phases. The path integral method (and the time-dependent Schrödinger equation) gives the exact phase difference in all situations. It is therefore an appropriate method to use in conceptual discussions of matter wave diffraction.

In some physics textbooks, both paths and waves are omitted from the description of matter wave diffraction. Instead, the discussion refers back to water waves or Young's experiment for light waves and quotes the condition for interference or phase differences by analogy.[24, 25, 26, 27, 28] This presentation is correct to obtain phase differences, but it ignores the differences in propagation, that is the time dependent behavior, between light and matter waves.

Some physics textbooks,[26, 27, 28] as well as some advanced undergraduate and graduate texts, will draw attention to group and phase velocities in sections unrelated to the double slit description.[16, 29, 30] It is interesting to contemplate at what level and in what manner the conceptual difficulty discussed in this paper could be addressed. For example, it could follow a discussion of the group and phase velocities of a matter wave packet. The results from the path integral formalism could thus be presented at the undergraduate level to elucidate the idea of a "path."[31, 32, 33]

**Appendix A: Time sum derivation**

In the following, the propagator is derived for a single slit crossing as described in Section IIIc. In Eq. ( 20 ), each of the positions $x_k$ represent the range of positions a point could have along a path at the k[th] time step of the sum. Requiring that a path intersects the slit at $x_{slit}$ in the k[th] time step is defined as a multiplication of a term $\delta(x_{slit} - x_k)$ to the integrand. This intersection happens at any time step from the first up to the last, so a factor

$$\chi_{slit} = \sum_{k=1}^{N-1} \delta(x_{slit} - x_k) \qquad (A\ 1)$$

must be included in Eq. ( 20 ) to describe all of the alternative times a path can intersect the slit. Substituting Eq. ( A 1 ) into Eq. ( 20 ) gives the total amplitude to travel from $\alpha$ to $\beta$ as

$$K(\beta;\alpha) = \lim_{\epsilon \to 0} \left(\frac{m}{2\pi i \hbar \epsilon}\right)^{N/2} \int \cdots \int \chi_{slit} \exp \frac{im}{2\hbar\epsilon} \left\{\sum_{j=1}^{N} (x_j - x_{j-1})^2\right\} dx_1 \ldots dx_{N-1}$$

$$= \lim_{\epsilon \to 0} \sum_{k=1}^{N-1} \left(\frac{m}{2\pi i \hbar \epsilon}\right)^{N/2} \int \cdots \int \exp \frac{im}{2\hbar\epsilon} \left\{\sum_{j=1}^{N} (x_j - x_{j-1})^2\right\} \delta(x_{slit} - x_k) dx_1 \ldots dx_{N-1}$$

(A 2)

Performing the $N - 1$ integrations in Eq. ( A 2 ) leads to the total sum

$$K(\beta;\alpha) = \lim_{\epsilon \to 0} \sum_{k=1}^{N-1} K(x_\beta, t_\beta; x_{slit}, t_\alpha + k\epsilon) \cdot K(x_{slit}, t_\alpha + k\epsilon; x_\alpha, t_\alpha). \qquad (A\ 3)$$

The substitution $t_{slit} \equiv t_\alpha + k\epsilon$ is made in Eq. ( A 3 ) to obtain the final result

$$K(\beta;\alpha) = \lim_{\epsilon \to 0} \sum_{t_{slit}=t_\alpha+\epsilon}^{t_\beta-\epsilon} K(x_\beta, t_\beta; x_{slit}, t_{slit}) \cdot K(x_{slit}, t_{slit}; x_\alpha, t_\alpha)$$

$$= \sum_{t_{slit}=t_\alpha}^{t_\beta} K(x_\beta, t_\beta; x_{slit}, t_{slit}) \cdot K(x_{slit}, t_{slit}; x_\alpha, t_\alpha).$$

(A4)

**Appendix B: Evaluating the integral of the full time sum**

The time sum derived in Eq. ( A 4 ) over the continuous value $t_{slit}$ must be handled carefully near the singular points occurring at $t_{slit} = t_\beta$ and $t_{slit} = t_\alpha = 0$, so we convert the sum to an integral prior to performing the limit $\epsilon \to 0$ to obtain

$$I(\beta; \alpha) = \lim_{\epsilon \to 0} \int_{\epsilon}^{t_\beta - \epsilon} dt_{slit} \frac{m}{2\pi i\hbar} \left(\frac{1}{t_{slit}(t_\beta - t_{slit})}\right)^{1/2} \exp\left\{\frac{imL^2 \cdot t_\beta}{2\hbar t_{slit}(t_\beta - t_{slit})}\right\}, \quad (\text{B 1})$$

where $(x_\beta - x_\alpha)/2 = L$ as before. Next, $x = 2(t_{slit}/t_\beta - 1/2)$ is substituted to obtain

$$I(\beta; \alpha)$$

$$= \lim_{\epsilon \to 0} \int_{-1+\frac{2\epsilon}{t_\beta}}^{1-\frac{2\epsilon}{t_\beta}} dx \frac{m}{2\pi i\hbar} \left(\frac{1}{(1+x)(1-x)}\right)^{1/2} \exp\left\{\frac{i2mL^2}{\hbar t_\beta} \frac{1}{(1+x)(1-x)}\right\}. \quad (\text{B 2})$$

Equation ( B 2 ) is simplified by the definition of the stationary phase as $\varphi_0 \equiv 2mL^2/\hbar t_\beta$ as in Section IIIc. The next substitution to be performed is $x = \sin(\theta)$. This trigonometric substitution eliminates the square root, as $dx/\sqrt{1-x^2} = d\theta$, and we obtain

$$I(\beta; \alpha) = \lim_{\epsilon \to 0} \frac{m}{2\pi i\hbar} \int_{\sin^{-1}\left(-1+\frac{2\epsilon}{t_\beta}\right)}^{\sin^{-1}\left(1-\frac{2\epsilon}{t_\beta}\right)} d\theta \, \exp\left(\frac{i\varphi_0}{\cos^2(\theta)}\right)$$

$$= \lim_{\epsilon \to 0} \frac{m}{2\pi i\hbar} \int_{\sin^{-1}\left(-1+\frac{2\epsilon}{t_\beta}\right)}^{\sin^{-1}\left(1-\frac{2\epsilon}{t_\beta}\right)} d\theta \, \exp(i\varphi_0) \cdot \exp(i\varphi_0 \tan^2(\theta)), \quad (\text{B 3})$$

where the identity $sec^2(\theta) = 1 + tan^2(\theta)$ is used in order to factor out a term $exp(i\varphi_0)$. Next, we substitute $u = tan(\theta)$ and obtain

$$I(\beta;\alpha) = \lim_{\epsilon \to 0} \frac{m}{2\pi i\hbar} e^{i\varphi_0} \int_{tan\left(sin^{-1}\left(-1+\frac{2\epsilon}{t_\beta}\right)\right)}^{tan\left(sin^{-1}\left(1-\frac{2\epsilon}{t_\beta}\right)\right)} du \, \frac{exp(i\varphi_0 u^2)}{1+u^2}. \quad (B\,4)$$

The limits of integration are symmetric and now tend to $\pm\infty$ as $\epsilon \to 0$, so they are redefined as $tan\left(sin^{-1}\left(1-\frac{2\epsilon}{t_\beta}\right)\right) = R$ and $tan\left(sin^{-1}\left(-1+\frac{2\epsilon}{t_\beta}\right)\right) = -R$, with the limit $R \to \infty$. Finally, we extend the integrand into the complex plane by performing the substitution $t = -i\sqrt{i\varphi_0}u$ to obtain

$$I(\beta;\alpha) = \lim_{R \to \infty} \frac{m}{2\pi i\hbar} e^{i\varphi_0} i\sqrt{i\varphi_0} \int_{-i\sqrt{i\varphi_0}(-R)}^{-i\sqrt{i\varphi_0}R} dt \, \frac{exp(-t^2)}{\left(\sqrt{i\varphi_0}\right)^2 - t^2}. \quad (B\,5)$$

The integrand is analytic everywhere in the complex plane except for first-order poles at $\pm\sqrt{i\varphi_0}$, therefore the path of integration, which lies on the line $t = R\, exp(i\,3\pi/4)$, can be rotated to lie entirely on the real axis. The integrand's even symmetry then permits

$$I(\beta;\alpha) = \frac{m}{2i\hbar} e^{i\varphi_0} \left[\frac{2i}{\pi} \sqrt{i\varphi_0} \int_0^\infty dt \, \frac{exp(-t^2)}{\left(\sqrt{i\varphi_0}\right)^2 - t^2}\right]. \quad (B\,6)$$

The term in square brackets of Eq. (B 6) has the form of the complex-valued function $w(z)$, given in Eq. 7.1.4 of Ref. 21, as

$$w(z) = \frac{2iz}{\pi} \int_0^\infty dt \, \frac{exp(-t^2)}{z^2 - t^2}. \quad (B\,7)$$

This function can be readily evaluated from the definitions given in Eqs. 7.1.2 and 7.1.3 of Ref. 21, as

$$w(z) = e^{-z^2} erfc(-iz), \quad (B\,8)$$

where $erfc(z)$ is the complementary error function. Substituting Eqs. ( B 7 ) and ( B 8 ) into Eq. ( B 6 ), we obtain the result,

$$I(\beta; \alpha) = \frac{m}{2\pi i \hbar} \pi \cdot erfc\left(-i\sqrt{i\varphi_0}\right), \quad (B\,9)$$

which is Eq. ( 26 ).

## VI. Acknowledgements

The authors gratefully acknowledge discussions with Ron Capelletti, Sam Werner, Mike Snow, Bradley Shadwick, and Ilya Fabrikant. E. Jones, R. Bach, and H. Batelaan gratefully acknowledge funding from the NSF (grant numbers 0969506 and 1306565). E. Jones also gratefully acknowledges funding from the DOE GAANN program number 58558.